\documentclass[12pt]{article}

\begin{document}

\newpage
\pagenumbering{arabic}

\begin{center}
 \bf \Large The $n$-wave procedure and dimensional regularization for
the scalar field in a homogeneous isotropic space
\end{center}

\begin{center}
{\bf    Yu.\,V. Pavlov{\,}\footnote{E-mail:\ pavlov@lpt.ipme.ru }}  \\[4mm]
{\small \it
  Institute of Mechanical Engineering, Russian Academy of Sciences, \\
  61 Bolshoy, V.O., St.\,Petersburg, 199178, Russia; \\[2mm]
A.\,Friedmann Laboratory for Theoretical Physics,\\
St.\,Petersburg, Russia
}
\end{center}

\begin{abstract}
\noindent
{\bf Abstract.}
    We obtain expressions for the vacuum expectations of
the energy-momentum tensor of the scalar field with an arbitrary
coupling to the curvature in an $N$-dimensional homogeneous isotropic
space for the vacuum determined by diagonalization of the Hamiltonian.
    We generalize the $n$-wave procedure to $N$-dimensional
homogeneous isotropic space-time.
    Using the dimensional regularization, we investigate the geometric
structure of the terms subtracted from the vacuum energy-momentum tensor
in accordance with the $n$-wave procedure.
   We show that the geometric structures of the first three
subtractions in the $n$-wave procedure and in the effective action method
coincide.
    We show that all the subtractions in the $n$-wave procedure
in a four- and five-dimensional homogeneous isotropic spaces correspond to
a renormalization of the coupling constants of the bare gravitational
Lagrangian.

\vspace{14pt}
\hfill 
\end{abstract}

\section{Introduction}
\hspace{\parindent}
    Quantum field theory in a curved space-time is intensively developed
since the 70s of the last century (see books~\cite{GMM}, \cite{BD})
and has important applications to the cosmology and astrophysics.
    In particular, the particle creation from the vacuum by
a gravitational field can be used to explain the observed density of
the visible and dark matter~\cite{PGr2000}.
    There exist models explaining the observed acceleration of
the expansion of the Universe by effects of quantum field theory in
a curved space~\cite{ParR01}.

    In investigating the effect of a quantized field on the space-time
metric, it is necessary to evaluate the vacuum expectations of
the energy-momentum tensor (EMT), which are sources of the
gravitational field in accordance with the semiclassical approach.
    Because the vacuum expectations of the EMT are represented
by divergent expressions, renormalization is required.
    There exist different methods for eliminating divergences of
the vacuum EMT (see the reviews in~\cite{GMM} and \cite{BD}).
    The $n$-wave procedure, proposed in~\cite{ZlSt}, and the equivalent
subtraction scheme~\cite{PaFu74}, called the adiabatic regularization,
are widely used for homogeneous isotropic spaces.
    They have been used in a great number of calculations
(see~\cite{GMM}, \cite{BD} for the scalar field with the conformal
coupling to curvature and see, e.g., \cite{BLM}, \cite{HMPM00}
for an arbitrary coupling).
    Rigorously justifying the expressions obtained for the vacuum EMT
requires interpreting the relevant subtraction schemes in terms
of renormalized constants of the bare Lagrangian.
    For the scalar field with an arbitrary coupling to curvature,
the geometric structure of infinite subtractions in the adiabatic
procedure in a homogeneous isotropic space was established
in~\cite{Bunch80}.
    But the geometric structure of finite counterterms
and their interpretation in terms of renormalization have remained
unexplored.
    For a space-time with flat spatial sections, the complete geometric
structure of the counterterms in the $n$-wave procedure was determined
using the dimensional regularization in~\cite{MMSH}, \cite{Pv2}.
    In~\cite{Pv2}, it was shown that all the subtractions in the $n$-wave
procedure in the four-dimensional spase-time with flat spatial sections
correspond to a renormalization of the constants of the bare
dimensionally regularized gravitational Lagrangian.

    This work is devoted to investigating the geometric structure of
counterterms in the $n$-wave procedure for nonflat spatial sections and
to comparing it with the results obtained from the dimensionally
regularized effective action.
    We note that calculations for the $N$-dimensional space-time
that are required for performing the dimensional regularization in this
work can also be of independent interest for investigating quantum theory
effects in curved space-time in higher-dimensional models.

     In this work, we consider the complex scalar field with an arbitrary
coupling to the curvature in the $N$-dimensional space-time with
homogeneous isotropic spatial sections.
    In Sec.~2, we describe the geometric structure of the expressions
subtracted from the vacuun EMT in the effective action method.
    In Sec.~3, we obtain the expressions for the vacuum expectations
of the EMT in the homogeneous isotropic space for the vacuum
defined in accordance with the Hamiltonian diagonalization method.
    In Sec.~4, we generalize the $n$-wave procedure for the $N$-dimensional
homogeneous isotropic case and give the first three counterterms,
which exhaust all counterterms in dimensions $N=4,5$.
    Using dimensional regularization, we investigate the geometric
structure of the $n$-wave procedure counterterms.
    In Sec.~5, the main results of the work are formulated.
    In Appendix~A, we give some formulas for eigenfunctions of
the Laplace-Beltrami operator in the $(N-1)$-dimensional space of constant
curvature used to calculate the vacuum expectations of the EMT.
    In Appendix~B, we give expressions for the geometric
quantities in the $N$-dimensional homogeneous isotropic space-time
encountered in this work.

     We use the system of units where $\hbar =c=1$.
     The signs of the curvature tensor and the Ricci tensor are
chosen such that
$ R^{\, i}_{\ jkl}= \partial_l \, \Gamma^{\, i}_{\, jk} -
\partial_k \, \Gamma^{\, i}_{\, jl} +
\Gamma^{\, i}_{\, nl} \Gamma^{\, n}_{\, jk} -
\Gamma^{\, i}_{\, nk} \Gamma^{\, n}_{\, jl}\ $
and  $\, R_{ik} = R^{\, l}_{\ ilk}$\,,
where $\Gamma^{\, i}_{\, jk}$ are Christoffel symbols.

\section{Geometric structure of counterterms in
         the effective action method}

\hspace{\parindent}
      We consider a complex scalar field $\varphi(x)$ of mass $m$
with the equation of motion
\begin{equation}
 ({\nabla}^i {\nabla}_{\! i} + \xi R +m^2) \varphi(x)=0 \, ,
\label{Eqm}
\end{equation}
     that corresponds to the Lagrangian
\begin{equation}
L(x)=\sqrt{|g|}\ [\,g^{ik}\partial_i\varphi^*\partial_k\varphi -
(m^2+\xi R)\, \varphi^* \varphi \,] \,,
\label{Lag}
\end{equation}
    where ${\nabla}_{\! i}$ are covariant derivatives corresponding
to the metric $g_{ik}$,
$\ g\!=\!{\rm det}(g_{ik})$,  $\ R$ is the scalar curvature,
and $\xi$ is the coupling constant to the curvature.
    In the space-time of dimension $N$ with
$ \xi =\xi_c \equiv (N-2)/\,[\,4\,(N-1)] $
(the conformal coupling) and $m=0$,
Eq.~(\ref{Eqm}) is conformally invariant \ ($\xi_c=1/6$ for $N=4$).

     The metric EMT, obtained by varying the action with respect to
the metric, is given by (see~\cite{ChT})
    \begin{equation}
T_{ik}= \partial_i\varphi^* \partial_k\varphi+ \partial_k\varphi^*
\partial_i\varphi
- g_{ik} |g|^{-1/2} L(x) - 2 \xi ( R_{ik}+\nabla_{\! i} \nabla_{\! k}
- g_{ik}\nabla^j \nabla_{\! j} ) \varphi^* \varphi .
\label{Tem}
\end{equation}
    Expressions for the vacuum expectations of the EMT are devergent.
    In a space-time with a metric of the general form, it is convenient
to analyze the geometric structure of the divergences using
the dimensionally regularised effective action.
    For the complex scalar field $\varphi(x)$
with the equation of motion~(\ref{Eqm}), the one-loop effective action
can be written as (see~\cite{BD}, \cite{Bunch79})
       \begin{equation}
S_{eff} = \int L_{eff}(x) \sqrt{|g|}\, d^N x \,,
\label{Seff}
\end{equation}
   where
       \begin{equation}
L_{eff}(x) = (4 \pi)^{-N/2} \left( \frac{M}{m} \right)^{2\varepsilon}
\sum_{j=0}^\infty a_j(x) \, m^{N_0-2j}\,
\Gamma\biggl(j-\frac{N}{2}\biggr)  \,,
\label{Leff}
\end{equation}
       \begin{equation}
a_0(x)=1 \,, \ \ \ \ \ a_1(x) = \biggl( \frac{1}{6} - \xi \biggr) \,R \,,
\label{a0a1}
\end{equation}
       \begin{equation}
a_2(x) = \frac{1}{180} R_{lmpq} R^{\,lmpq} -
\frac{1}{180} R_{lm} R^{\,lm} -
\frac{1}{6} \biggl( \frac{1}{5} - \xi \biggr) \nabla^l \nabla_{\!l} R
\, + \frac{1}{2} \biggl( \frac{1}{6} - \xi \biggr)^2 R^2,
\label{a2}
\end{equation}
    $N$ is the space-time dimension, which is considered as variable
analytically continued into the complex plane,
$\varepsilon$ is a complex parameter,
$M$ is a constant with the dimension of mass~\cite{tHooft} introduced
to preserve the standard dimension (length)${}^{-N_0}$ of $L_{eff}$ in
the case $N=N_0-2\varepsilon$, and $ \Gamma(z)$ is the Gamma function.

    The first $[N_0/2]+1$ terms in~(\ref{Leff})
are to be eliminated to obtain the renormalized Lagrangian $L_{eff}$
($[b]$ denotes the integer part of $b$).
   Varying the effective action terms corresponding to $j=0,1$ with
respect to $g_{ik}$, we obtain the terms subtracted from the vacuum EMT,
     \begin{equation}
T_{ik,\varepsilon}[0]=- \frac{m^{N_0}}{2^{N_0} \pi^{N_0/2}}
\left( \frac{4 \pi M^2}{m^2} \right)^{\! \displaystyle \varepsilon }
\Gamma  \biggl( \varepsilon - \frac{N_0}{2} \biggr) \, g_{ik} \,,
\label{TE0}
\end{equation}
    \begin{eqnarray}
T_{ik,\varepsilon}[1] = \frac{m^{N_0-2}}{2^{N_0-1} \pi^{N_0/2}}
\left( \frac{1}{6} - \xi \right)
\left(\frac{4 \pi M^2}{m^2} \right)^{\! \displaystyle \varepsilon }
\Gamma \biggl( 1 -  \frac{N}{2} \biggr) \, G_{ik} =
\nonumber     \\
= \frac{m^{N_0-2}}{2^{N_0-1} \pi^{N_0/2}}\!
\left(\frac{4 \pi M^2}{m^2} \right)^{\! \displaystyle \varepsilon } \!
\Biggl[\frac{-\, \Gamma \left(3 - \frac{N}{2} \right)}
{3 ( N - 1 ) ( N - 2 )} +
\Delta \xi\, \Gamma \biggl(1 - \frac{N}{2} \biggr) \Biggr]  G_{ik}\,,
\label{TE1}
\end{eqnarray}
    where $G_{ik}= R_{ik} - R g_{ik}/2$ is the Einstein tensor and
$  \Delta \xi \equiv \xi_c - \xi $.

    We find the term $T_{ik,\varepsilon}[2]$, corresponding to
subtraction of the term with $j=2$ from $L_{eff}$.
    To analysis the case of homogeneous isotropic spaces,
it is convenient to rewrite~(\ref{a2}) as
    \begin{eqnarray}
a_2(x) =\! \frac{N-6}{720 (N\!-\!3)} (R_{lmpq} R^{\,lmpq} \!-\!
4 R_{lm} R^{\,lm} \!+\! R^2) +
\frac{(N\!\!-\!2) C_{lmpq} C^{\,lmpq} }{240 (N - 3)}  -
\nonumber        \\
-\, \frac{1}{6} \biggl( \frac{1}{5} - \xi \biggr) \nabla^l \nabla_{\!l} R
+ \left( \frac{(N\!-\!4) (N\!-\!6)}{480 (N-1)^2}
+\frac{\Delta \xi (4\!-\!N)}{12(N\!-\!1)} +\frac{(\Delta \xi)^2}{2}
 \right) R^2 \,, \phantom{x}
\label{a2m}
\end{eqnarray}
     where
\begin{equation}
C_{iklm} = R_{iklm} + \frac{2}{N\!-\!2} \biggl( R_{m\,[\,i}\, g_{k\,]\,l} -
R_{l\,[\,i}\, g_{k\,]\,m} \biggr) +
\frac{2 \, R \,g_{l\,[\,i}\,g_{k\,]\,m} }{(N\!-\!1)(N\!-\!2)}
\label{Ciklm}
\end{equation}
    is the conformal Weyl tensor and
    the square brackets in subscripts denote antisymmetrization:
$A_{n[i} B_{k]m}=(A_{ni} B_{km}-A_{nk} B_{im})/2$.
    Using~(\ref{a2m}) and varying the effective action term corresponding
to $j=2$ with respect to $g_{ik}$, we obtain
$T_{ik,\varepsilon}[2]$ in the form
      \begin{eqnarray}
T_{ik,\varepsilon}[2] \!=\! \frac{m^{N_0-4}}{(4\pi)^{\frac{N_0}{2}}}
\!\left( \!\frac{4 \pi M^2}{m^2}\!\right)^{\! \displaystyle \varepsilon }
\!\left[ \!\frac{(N\!-\!6) E_{ik}}{360 (N\!-\!3)}
\Gamma \biggl(2 \!-\! \frac{N}{2}\biggr) +
\frac{(N\!-\!2) W_{ik}}{120 (N\!-\!3)} \Gamma \biggl(2 \!-\!
\frac{N}{2}\biggr) \! \right.\! +
\hspace{-4mm} \nonumber                \\
+\left.\Biggl( \, \frac{\Gamma \left( 4 - \frac{N}{2}\right)}{60 \,
(N - 1)^2 }
+ \Delta \xi\, \frac{\Gamma \left(3 - \frac{N}{2} \right)}{3\, (N - 1) }
+ (\Delta \xi)^2  \, \Gamma \biggl( 2 - \frac{N}{2} \biggr) \,\Biggr)
{}^{(1)}\! H_{ik} \right]\!, \phantom{xx}
\label{TE2ik}
\end{eqnarray}
    where
    \begin{equation}
{}^{(1)}\! H_{ik}= \frac{\delta  {\displaystyle \int \!  R^2  \,
\sqrt{|g|} \, d^N x}} {\sqrt{|g|} \,  \delta g^{ik}} =
2 \biggl( \nabla_{\! i} \nabla_{\! k} R - g_{ik} \nabla^l \nabla_{\! l} R
\biggr) + 2 R \biggl( R_{ik} -\frac{1}{4} R \, g_{ik} \biggr),
\label{1Hik}
\end{equation}
    \begin{equation}
E_{ik} =  \frac{1}{\sqrt{|g|}} \, \frac{\delta}{\delta g^{ik}}
\int \! (R_{lmpq} R^{\,lmpq} - 4 R_{lm} R^{\,lm} + R^2)
 \, \sqrt{|g|} \,d^N x \,,
\label{Eik}
\end{equation}
    \begin{equation}
W_{ik} =  \frac{1}{\sqrt{|g|}} \, \frac{\delta}{\delta g^{ik}}
\int \! C_{lmpq} C^{\,lmpq} \, \sqrt{|g|} \,d^N x \,.
\label{Wik}
\end{equation}
    In the four-dimensional space-time, $E_{ik}=0$ because
     \begin{equation}
\frac{\delta}{\delta g^{ik}} \int \! \left( R^{\,lmpq} R_{lmpq} -
4 R^{\,lm} R_{\,lm} + R^2 \right) \sqrt{|g|} \,d^4 x = 0
\label{GaussB}
\end{equation}
    in accordance with the Gauss-Bonnet theorem.
    Using (\ref{Ciklm}), (\ref{1Hik}) and the formulas
      \begin{eqnarray}
{}^{(2)}\! H_{ik}= \frac{1}{\sqrt{|g|}} \, \frac{\delta}{\delta g^{ik}}
\int \! R^{\,lm} R_{lm} \, \sqrt{|g|} \,d^N x  =
\nabla_{\! i} \nabla_{\! k} R -  \nabla^l \nabla_{\! l} R_{ik} -
\nonumber           \\
-\, \frac{1}{2} \left( \nabla^l \nabla_{\! l} R +R^{\,lm}R_{lm} \right)
g_{ik} + 2 R^{\,lm} R_{limk} \,,    \phantom{xxxxxxxxx}
\label{2Hik}
\end{eqnarray}
       \begin{eqnarray}
H_{ik}= \frac{1}{\sqrt{|g|}} \, \frac{\delta}{\delta g^{ik}}
\int \! R^{\,lmpq} R_{lmpq} \, \sqrt{|g|} \,d^N x  =
2 \nabla_{\! i} \nabla_{\! k} R - 4 \nabla^l \nabla_{\! l} \, R_{ik} -
\nonumber           \\
- \, 4 R_{il}  R_k^{\, l} + 4 R^{\,lm} R_{limk}
- \frac{1}{2} \, g_{ik} R_{lmpq} R^{\, lmpq} +
2 R_{ilmp} R_k^{\ lmp}
\label{Hik}
\end{eqnarray}
    for variational derivatives of the expressions that are quadratic in
the curvature (see~\cite{DeWitt}), we can obtain (see~\cite{Bunch79})
       \begin{equation}
E_{ik} \!=\! H_{ik} - 4{}^{(2)}\!H_{ik} + {}^{(1)}\!H_{ik} \!=\!
2 C_{ilmp} C_k^{\ lmp} -\frac{g_{ik}}{2} C_{lmpq} C^{lmpq} -
(N\!-\!4) {}^{(3)}\!H_{ik}
\label{ECH}
\end{equation}
for arbitrary $N$, where
      \begin{eqnarray}
{}^{(3)}\!H_{ik} = \frac{4}{N-2} \,C_{ilkm} R^{\,lm} +
\frac{2(N-3)}{(N-2)^2} \left[ 2 R_{il} R_k^{\,l} -
\frac{N}{N-1}\, R R_{ik}  \right. -   \phantom{xxxx} \nonumber    \\
- \, \left. g_{ik} \biggl( R_{lm} R^{\,lm} -
\frac{N+2}{4(N-1)}\,R^2 \,\biggr) \right]. \phantom{xxxx}
\label{3Hik}
\end{eqnarray}
    In the conformally flat case~($ C_{iklm} = 0 $),
the tensor $ {}^{(3)}\!H_{ik} $ is covariantly conserved,
   which follows from (\ref{ECH}) for $N\ne4$
and can be directly verified for $N=4$.
   The tensor $ {}^{(3)}\!H_{ik} $  proposed in~\cite{Bunch79}
generalizes the corresponding tensor introduced in~\cite{GKL71}
for the conformally flat four-dimensional case.

    As $N\to 4$, the product $E_{ik}\Gamma (2 - (N/2))$
has a finite limit for an arbitrary space-time metric
because in the analitic continuation with respect to the dimension,
the dependence of the expressions on $N$ is assumed to be given by
a rational function, $E_{ik}=0$ at $N=4$, and the Gamma function
has a first-order pole at zero.
    Therefore, the first term in the square brackets in the right-hand
side of~(\ref{TE2ik}) is always finite as $N\to 4$.
    If this term is not subtracted in renormalization the EMT, then
the vacuum EMT is finite as $N \to 4$,
but the effective action then remains divergent, and the expression for
the anomalous trace of the vacuum EMT is differnt from the standard one.
    Under an additional finite renormalization of the coefficient
at $R^2$ in $L_{eff}$
(see a discussion of this point in Sec.~6.3 in~\cite{BD}),
the anomalous trace can be made vanishing for the conformal scalar field
in the conformally flat space.
    This occurs, e.g., in~\cite{MMSt}, where the time-dependent normal
ordering of operators was used to obtain finite values of the vacuum EMT
in a homogeneous isotropic space for the conformal scalar field.

    In what follows, in accordance with standard approach, we keep
the term with $E_{ik}$ in~(\ref{TE2ik}).
    In the conformally flat case, in particular, in a homogeneous isotropic
space-time, taking the equalities $C_{iklm}=0, W_{ik}=0$ and (\ref{ECH})
into account, we obtain the following expression for the term
$T_{ik,\varepsilon}[2]$ that is
subtracted from the vacuum EMT in the effective action method:
      \begin{eqnarray}
T_{ik,\varepsilon}[2] = \frac{m^{N_0-4}}{(4\pi)^{N_0/2}}
\left( \frac{4 \pi M^2}{m^2}\right)^{\! \displaystyle \varepsilon }
\left[ \frac{-{}^{(3)}\! H_{ik}}{90 (N-3)}\,
\Gamma \biggl(4 - \frac{N}{2}\biggr) \right. \! +
     \nonumber                \\
+\left.\Biggl( \, \frac{\Gamma \left( 4 - \frac{N}{2}\right)}{60 \,
(N - 1)^2 }
+ \Delta \xi\, \frac{\Gamma \left(3 - \frac{N}{2} \right)}{3\, (N - 1) }
+ (\Delta \xi)^2  \, \Gamma \biggl( 2 - \frac{N}{2} \biggr) \,\Biggr)
{}^{(1)}\! H_{ik} \right]. \phantom{x}
\label{T2ikc}
\end{eqnarray}

\section{Scalar field in the homogeneous isotropic space}
\hspace{\parindent}
    The metric of the $N$-dimensional homogeneous isotropic space can be
written as
    \begin{equation}
ds^2=g_{ik}\,dx^i dx^k = a^2(\eta)\,(d{\eta}^2 - d l^2) \,,
\label{gik}
\end{equation}
    where $ d l^2=\gamma_{\alpha \beta}\, d x^\alpha d x^\beta $
is the metric of the $(N-1)$-dimensional space of constant curvature
$K=0, \pm 1 $.
    The complete set of solutions of Eq.~(\ref{Eqm}) in metric~(\ref{gik})
can be found in the form
    \begin{equation}
\varphi(x) = a^{-(N-2)/2} (\eta)\, g_\lambda (\eta)
\Phi_J ({\bf x}) \,,
\label{fgf}
\end{equation}
    where
\begin{equation}
g_\lambda''(\eta)+\Omega^2(\eta)\,g_\lambda(\eta)=0 \,,
\label{gdd}
\end{equation}
       \begin{equation}
\Omega^2(\eta)=m^2 a^2 +\lambda^2 - \Delta \xi\,a^2 R\,,
\label{Ome}
\end{equation}
     \begin{equation}
\Delta_{N-1}\,\Phi_J ({\bf x}) = - \left( \lambda^2 -
\left(\frac{N-2}{2} \right)^2 K \right) \Phi_J  ({\bf x})\,,
\label{DFlF}
\end{equation}
    the prime denotes the derivative with respect to the conformal
time~$\eta$,
  and $J$ is a set of indices (quantum numbers)  that label
the eigenfunctions of the Laplace-Beltrami operator $\Delta_{N-1}$
in the ($N-1$)-dimensional space.
    The nonnegativity of the eigenvalues of the operator $-\Delta_{N-1} $
implies the inequality $ \lambda^2-((N-2)/\,2)^2\,K \ge 0 $\,.

    In according with the Hamiltonian diagonalization method~\cite{GMM}
(see~\cite{Pv} for the case of an arbitrary $\xi$),
the function $g_\lambda(\eta)$ must satisfy the initial conditions
    \begin{equation}
g_\lambda'(\eta_0)=i\, \Omega(\eta_0)\, g_\lambda(\eta_0) \,, \ \ \
|g_\lambda(\eta_0)|= \frac{1}{ \sqrt{\Omega(\eta_0)} } \,.
\label{icg}
\end{equation}

     For quantizing, we expand the field $ \varphi(x) $ with respect to
the complete set of solutions of Eq.~(\ref{fgf})
\begin{equation}
\varphi(x)=\int \! d\mu(J)\,\biggl[ \varphi^{(+)}_J \,a^{(+)}_J +
\varphi^{(-)}_J \, a^{(-)}_J \,\biggr] \ ,
\label{fff}
\end{equation}
      where
\begin{equation}
\varphi^{(+)}_J (x)=\frac{a^{-(N-2)/2} (\eta)}{\sqrt{2}}\,
g_\lambda(\eta)\,\Phi^*_J({\bf x}) \ ,
\ \  \ \
\varphi^{(-)}_J (x)=\biggl(\varphi_J^{(+)}(x) \biggr)^* \ ,
\label{fpm}
\end{equation}
     and impose the commutation relations
\begin{equation}
\left[a_J^{(-)}, \ \stackrel{*}{a}\!{\!}_{J'}^{(+)}\right]=
\left[\stackrel{*}{a}\!{\!}_J^{(-)}, \ a_{J'}^{(+)}\right]=\delta_{JJ'} \ ,
\ \ \ \left[a_J^{(\pm)}, \ a_{J'}^{(\pm)}\right]=
\left[\stackrel{*}{a}\!{\!}_J^{(\pm)}, \ \stackrel{*}{a}\!{\!}_{J'}^{(\pm)}
\right]=0 \,.
\label{aar}
\end{equation}
     It is convenient to express the EMT expectation values for
the vacuum $| 0\rangle $  annihilated by the operators
$a_J^{(-)}\!,\ \stackrel{*}{a}\!{\!}_{J}^{(-)}$
in terms of the bilinear combinations of
the functions~$g_\lambda $ and~$g_\lambda^*$,
    \begin{equation}
S=\frac{|g_\lambda'|^2 + \Omega^2\,|g_\lambda|^2}{4 \, \Omega}
-\frac{1}{2} \,, \ \
U=\frac{\Omega^2 \, |g_\lambda|^2- |g_\lambda'|^2}{2\, \Omega} \,  \,, \ \
V= - \frac{d (g_\lambda^* g_\lambda)}{2\, d \eta} \,,
\label{SUV}
\end{equation}
     which in accordance with (\ref{gdd}) satisfy the system of
differential equations
    \begin{equation}
S'= \frac{\Omega'}{2\, \Omega} \, U \ , \ \ \
U'= \frac{\Omega'}{\Omega} \, (1+ 2 S) - 2 \, \Omega V \ , \ \ \
V'= 2 \,\Omega \, U \,.
\label{sdu}
\end{equation}
      Recalling the initial conditions
$ S(\eta_0)=U(\eta_0)=V(\eta_0)=0 $  following from~(\ref{icg}),
the Eqs.~(\ref{sdu}) can be rewritten as a system of the Volterra
integral equations
     \begin{equation}
U(\eta) + i V(\eta) = \int_{\eta_0}^{\, \eta} \! w(\eta_1)\,
(1+2 S(\eta_1))\, \exp[ 2\,i\,\Theta(\eta_1,\eta)]\,d\eta_1  \,,
\label{ie1}
\end{equation}
\begin{equation}
S(\eta)=\frac{1}{2}\,\int_{\eta_0}^{\, \eta} \! d\eta_1 \,
w(\eta_1)\, \int_{\eta_0}^{\, \eta_1} \! w(\eta_2)\,
(1+2 S(\eta_2)) \cos[2\,\Theta(\eta_2,\eta_1)]\, d\eta_2 \,,
\label{ie2}
\end{equation}
      where
$$
w(\eta) =\frac{\Omega'(\eta)}{\Omega(\eta)} \ , \ \ \ \ \
\Theta(\eta_1, \eta_2) = \int_{\eta_1}^{\eta_2} \Omega(\eta)\,d\eta \,.
$$

     Substituting expression (\ref{fff}) in (\ref{Tem}) and using
(\ref{aar}), (\ref{SUV}) and formulas (\ref{fF})--(\ref{nfF})
(see Appendix~A), we obtain the (divergent) expressions for the vacuum
expectations of the EMT,
    \begin{equation}
\langle 0 |\,T_{ik}| 0\rangle =\frac{B_N}{a^{N-2}} \int
\! \tau_{ik} \, d\mu(\lambda) \,,
\label{Ttik}
\end{equation}
      where
$ B_N\!=\!\left[ 2^{N-3}\pi^{(N-1)/\,2} \, \Gamma((N\!-\!1)/2)
\right]^{-1}$
and ``spectral densities'' $\tau_{ik}$\ are given by
    \begin{equation}
\tau_{00}=\Omega \biggl( S+\frac{1}{2}\biggr) +
\Delta \xi \,(N\!-\! 1) \left[ c V + \biggl(c'+(N\!-\! 2) c^2\biggr)
\frac{1}{\Omega} \! \left(\!S+\frac{1}{2}  U +
\frac{1}{2}\right)\right] ,
\label{ts00}
\end{equation}
                      \begin{eqnarray}
\tau_{\alpha \beta} = \gamma_{\alpha \beta}
\left\{\frac{\lambda^2}{(N-1)\,\Omega} \left( S+\frac{1}{2} \right)-
\frac{\Omega^2-\lambda^2}{2 \,(N-1)\,\Omega} \, U  \right. -
\phantom{xxxx}
\label{tsab}   \\
-\, \Delta \xi \left[ \biggl((N\!-\!1)\,c'+ (N\!-\!2)\, K \biggr)
\frac{1}{\Omega}
\left. \left( S+\frac{1}{2}\, U +\frac{1}{2} \right) +
2\,\Omega \,U\! -\! (N\!-\!1)\,c\,V \right] \right\},
\hspace{-4mm}\nonumber
\end{eqnarray}
     and $\ c\equiv \! a'/a$. \
     The integration measure in~(\ref{Ttik}) in the four-dimensional
space-time is (see~\cite{GMM})
    \begin{equation}
\int\!d\mu(\lambda)\ldots = \left\{
\begin{array}{ll}
{\ \displaystyle
\int_0^\infty d \lambda\, \lambda^{2} } \ldots, \ \
&  K=0, -1 \,,   \\[14pt]
{\displaystyle
\sum \limits_{\lambda=1}^\infty \lambda^2 } \ldots,  &  K=1.
\end{array}             \right.
\label{mera}
\end{equation}
    In the $N$-dimensional case,
$$
\int d\mu(\lambda) \ldots =
\int d\lambda \, f(\lambda) 2 B_N^{-1}\ldots \,,
$$
    where the function $f(\lambda)$ is defined in~(\ref{fF}) and
can be written in the respective forms (\ref{flK0}) and (\ref{flK1})
in quasi-Euclidean $(K=0)$ and spherical $(K=1)$ cases.

    We note  that the expressions for the vacuum expectation of the EMT
given in~\cite{BLM} are valid only for $N=4$,
whereas they are evaluated in~\cite{GMM}
for the  vacuum state that does not correspond to
the Hamiltonian diagonalization method
(for $\xi \ne \xi_c $  and  $R \ne 0 $).

\section{The $n$-wave procedure}

\hspace{\parindent}
    The $n$-wave prosedure proposed in~\cite{ZlSt} is often used
to calculate the renormalized vacuum EMT in homogeneous isotropic spaces.
    In the quasi-Euclidean $(K=0)$ case, the procedure is as follows.
    If the vacuum expectations of some operator $A$ bilinear in the field
are expressed as
     \begin{equation}
\langle 0|A\,| 0\rangle = \int a(\lambda,m) \, d \lambda \,,
\label{intA}
\end{equation}
    the corresponding renormalized value is then obtained by
regularizing the contribution of each mode
(i.e., of the integrand with a given $\lambda$).
    For this, the replacements
$\lambda \to n \lambda $ and $ m \to n m $ are made,
and the terms of the $1/n$-asymptotic expansion $(n \to \infty)$
that make integral~(\ref{intA}) divergent are subtracted from
$a(\lambda, m)$.
    This procedure corresponds to subtraction of the contribution
of the so-called $n$-waves
(see~\cite{ZlSt} and Sec.~2.2 in~\cite{GMM} for more details).
    This method for eliminating divergences is equivalent to the
adiabatic regularization method based on introducing
the adiabaticity parameter for the metric variation and
on subtracting the first terms of the asymptotic expansion  of
$a(\lambda,m)$ in inverse powers of this parameter
in~(\ref{intA}) (see~\cite{BD}).

    In applying the $n$-wave procedure in the case of a spherical $(K=1)$
and a hyperbolic $(K=-1)$\,  $N$-dimensional space-time,
it is necessery to take into account the difference of the measure
in~(\ref{Ttik}) from the quasi-Euclidean case.
    Writing the integration measure as
$d\mu(\lambda)=\sigma(\lambda)\, d\lambda$,
we take the generalization of the $n$-wave procedure to the
$N$-dimensional homogeneous isotropic space-time to be given by
the expression
     \begin{equation}
\langle 0|\,T_{ik}| 0\rangle_{ren} = \frac{B_N}{a^{N-2}}
\lim_{\Lambda \to \infty} \Biggl[\, \int^{\Lambda} \!
\tau_{ik} \, d\mu(\lambda) -
\sum \limits_{l=0}^{[N/2]} \int_0^{\Lambda} \!
\lambda^{N-2} a_{ik}[l] \, d \lambda \Biggr] \,,
\label{Trik}
\end{equation}
      where
\begin{equation}
a_{ik}[l] = \frac{1}{l!} \lim_{n \to \infty}
\frac{\partial^{\, l}}{\partial (n^{-2})^l}
\left( \frac{ \tau_{ik} (n \lambda, n m)\, \sigma(n \lambda)}{n^{N-1}
\lambda^{N-2}}  \right).
\label{aikl}
\end{equation}
    In the four-dimensional case, this generalization coincides with
the standard definition of the $n$-wave procedure.
      The function~$\sigma(\lambda)$ accounts for the difference
of the integration measure from $\lambda^{N-2}\, d \lambda \, $ for
$K \ne 0$ and $N\ne 4$.

     We now find the explicit form of $a_{ik}[l]$.
     For this, we expand $S, U$, and $V$ (see~(\ref{SUV})) in inverse
powers of $n$ and replace
$\lambda \to n \lambda $  and  $ m \to n m $  with $ n \to \infty $:
$$
S = \sum_{k=1}^\infty n^{-k} S_k\,, \ \ \ \ldots
$$
     Using consecutive iterations in integral
equations~(\ref{ie1}), (\ref{ie2}) and the stationary phase method,
we obtain the first nonzero terms of the expansions,
     \begin{equation}
V_1=W \,,\ \ U_2=DW \,,\ \
S_2=\frac{1}{4} W^2 \,, \ \
V_3=\frac{1}{2} W^3 - D^2 W - \frac{\omega}{2} D\left(\frac{q}{\omega^3}
\right) \,,
\label{V1U2S2}
\end{equation}
           \begin{equation}
U_4=\frac{3}{2} \, W^2 DW - D^3 W - D \left( \frac{\omega}{2} \, D \biggl(
\frac{q}{\omega^3} \biggr) \right) + \frac{q}{2\omega^2} DW \ ,
\label{U4}
\end{equation}
           \begin{equation}
S_4=\frac{3}{16} \, W^4 + \frac{1}{4}\,(D W)^2 - \frac{1}{2}\, W D^2 W
-\frac{1}{4}\,\omega W D\biggl(\frac{q}{\omega^3}\biggr) \ ,
\label{S4}
\end{equation}
    where
\begin{equation}
q=\Delta \xi \, a^2 R \,, \ \ \omega=(m^2a^2+\lambda^2)^{1/2} \,, \ \ \
W=\frac{\omega\,'}{2\,\omega^2} \ , \ \ \
D=\frac{1}{2\,\omega} \, \frac{d}{d\eta} \ .
\label{qoWD}
\end{equation}

    It must be note that in calculating the quantities
$S_k, U_k$, and $V_k$, as in~\cite{GMM}, \cite{ZlSt}, \cite{MMSH},
we neglect the terms depending on the initial time instant $\eta_0$,
i.e., restrict ourself to the case of counterterms that are local in
the time $\eta$.
   In particular, nonlocal terms are absent whenever it is assumed that
the first $2[N/2]$ derivatives of the scale factor $a(\eta)$
of the metric vanish at the initial time instant.

    We use~(\ref{ts00}), (\ref{tsab}), (\ref{V1U2S2})--(\ref{S4})
and in what follows set the measure $\sigma(\lambda)\, d \lambda $
equal to
     \begin{equation}
\sigma(\lambda)= \lambda^{N-2} + \alpha_N \lambda^{N-4} +
\beta_N \lambda^{N-6} + \ldots \,,
\label{sig}
\end{equation}
      where $\alpha_N,  \beta_N, \ldots $ are the functions of $N$ and $K$,
which vanish at $N=4$, to be determined below.
   As a result, we obtain the expressions for $a_{ik}[l]$
     \begin{equation}
a_{00}[0] = \tau_{00}[0] = \frac{\omega}{2} \,, \ \ \ \
a_{\alpha \beta}[0] = \tau_{\alpha \beta}[0] = \gamma_{\alpha \beta}\,
\frac{\lambda^2}{2 (N-1) \, \omega} \ ,
\label{at00}
\end{equation}
    \begin{equation}
a_{ik}[1]=\tau_{ik}[1] + \frac{\alpha_N}{\lambda^2} \tau_{ik}[0]
\,, \ \ \ \
a_{ik}[2]=\tau_{ik}[2] + \frac{\alpha_N}{\lambda^2} \tau_{ik}[1]
+ \frac{\beta_N}{\lambda^4} \tau_{ik}[0] \,,
\label{at12}
\end{equation}
                   where
              \begin{equation}
\tau_{00}[1]=\omega \, S_2 + \Delta \xi\,(N-1) \, \biggl[\,c\,V_1 +
\frac{N-2}{4 \, \omega} \,(c^2-K) \,\biggr] \ ,
\label{t001}
\end{equation}
                      \begin{eqnarray}
\tau_{\alpha \beta}[1] = \frac{\gamma_{\alpha \beta}}{N\!-\!1}\,
\Biggl[ \, \frac{\lambda^2}{\omega} S_2 -
\frac{m^2 a^2}{2 \,\omega} \, \biggl(U_2 +\frac{q}{2\,\omega^2}
\biggr) \, \Biggr] +   \gamma_{\alpha \beta}
 \Delta \xi \, \Biggl[ (N\!-\!1)\, c\,V_1 -     \nonumber     \\
- \, 2\omega \biggl(U_2 +\! \frac{q}{2\,\omega^2} \biggr) +
\frac{1}{4\,\omega} \biggl(-4 \xi a^2 R +
(N\!-\!2)\left(c^2 (N\!-\!1) + K (N\!-\!3) \right) \biggr) \, \Biggr]  \,,
\label{tab1}
\end{eqnarray}
                      \begin{eqnarray}
&& \phantom{xxxxxxx}   \tau_{00}[2] = \omega\,\biggl(S_4+
\frac{q}{4\,\omega^2}\,U_2 + \frac{q^2}{16\, \omega^4} \biggr) +
    \label{t002}  \\
&& \mbox{+} \Delta \xi (N-1) \left[\, c\,V_3 +\frac{N-2}{2\,\omega}\,
(c^2-K) \, \biggl(S_2+\frac{1}{2}\,U_2+\frac{q}{4\,\omega^2}\biggr)
\,\right] \,,
\nonumber
\end{eqnarray}
                      \begin{eqnarray}
\tau_{\alpha \beta}[2] = \gamma_{\alpha \beta}\left\{\frac{1}{N-1}\,
\left[\frac{\lambda^2}{\omega}\,\biggl( S_4 +
\frac{q}{4\,\omega^2}\,U_2 + \frac{q^2}{16\, \omega^4} \biggr)
-\frac{m^2 a^2}{2 \,\omega} \, \biggl(U_4 +\frac{q^2}{4\,\omega^4}
\right. \right. +     \hspace{-4mm}  \nonumber   \\
\mbox{+}\, \frac{q}{\omega^2}\,S_2 \biggr) \, \Biggr]
\, + \, \Delta \xi \left[\,(N-1)\, c\,V_3 -2\,\omega \,
\biggl(U_4 + \frac{q^2}{4\,\omega^4}
+\frac{q}{\omega^2}\,S_2 \biggr)
   \right. + \phantom{xxx}  \label{tab2}       \\
+\, \frac{1}{2\omega} \biggl( -4 \xi a^2R+
(N\!-\!2) \left(c^2(N\!-\!1)+K(N\!-\!3) \right)\, \biggr)
\,\biggl( S_2 + \frac{1}{2}U_2 +  \frac{q}{4\omega^2} \biggr) \,
\Biggr] \, \Biggr\} \,.  \hspace{-4mm}    \nonumber
\end{eqnarray}
    These expressions exhaust all the subtractions in dimensions $N=4,5$.
    Additional counterterms occur for $ N\ge 6 $
(they are given in~\cite{Pv2} for the conformal scalar field with $K=0$).

    The vacuum EMT renormalized in accordance with~(\ref{Trik})  is
covariantly conserved.
     This follows from Eqs.~(\ref{at00}), (\ref{at12}) and the equalities
$\nabla^i (\tau_{ik}/a^{N-2})=0 $ and $\nabla^i (\tau_{ik}[l]/a^{N-2})=0 $,
which can be verified using~(\ref{sdu}), (\ref{ts00}), (\ref{tsab}),
(\ref{t001})--(\ref{tab2}).

    To find the geometric structure of the $n$-wave procedure
counterterms, we use dimensional regularization, setting,
as in Sec.~2, \, $N=N_0-2\varepsilon$, where $\varepsilon$
is a complex parameter.
    In calculating integrals in the dimensionally regularized counterterms
    \begin{equation}
T_{ik, \varepsilon }[l]=\frac{B_N}{a^{N-2}} (M)^{2 \varepsilon}
\int_{0}^{\infty} \! \lambda^{N-2} a_{ik, \varepsilon}[l] \, d\lambda \,,
\label{kTik}
\end{equation}
    where $a_{ik,\varepsilon}[l] $ are defined as
in (\ref{at00})--(\ref{tab2}) with replacement $N \to N_0 - 2\varepsilon $,
we use the equality
    \begin{equation}
\int_0^\infty  x^k\, (1+x^2)^{-p}\, dx =\frac{\Gamma\,(
\frac{k+1}{2})\,\Gamma\,(p-\frac{k+1}{2})}{2\, \Gamma\,(p)}
\label{iGf}
\end{equation}
    and the analitic continuation of the expression in its right-hand side.
    The result for the zeroth counterterm coincides with~(\ref{TE0}).
    Thus, both in the effective action method and in
the $n$-wave procedure, the zeroth subtraction corresponds to
renormalization of the cosmological constant.

    With formulas~(\ref{G00Gab}), (\ref{c1H00})--(\ref{c3Hab})
(see Appendix~B) taken into account,
the structure of the first and the second subtractions in the $n$-wave
procedure in a homogeneous isotropic space-time can be obtained by the
respective additions of
$\Delta T_{ik,\varepsilon}[1]$ and $\Delta T_{ik,\varepsilon}[2]$
to~(\ref{TE1}) and (\ref{T2ikc}), where
    \begin{equation}
\Delta T_{ik,\varepsilon}[1] = \frac{m^{N_0-2}}{2^{N_0-1} \pi^{N_0/2}}
\left(\frac{4 \pi M^2}{m^2} \right)^{\! \displaystyle \varepsilon }
D_{ik}[1] \,,
\label{DTE1}
\end{equation}
      \begin{equation}
\Delta T_{ik,\varepsilon}[2] = \frac{m^{N_0-4}}{(4\pi)^{N_0/2}}
\left( \frac{4 \pi M^2}{m^2}\right)^{\! \displaystyle \varepsilon }
D_{ik}[2]    \,,
\label{DT2c}
\end{equation}
    \begin{equation}
D_{00}[1]=-\frac{1}{6} \Gamma \left(3 - \frac{N}{2} \right)
\left( K + \frac{24 \, \alpha_N}{(N-2)(N-3)(N-4)} \right),
\label{D100}
\end{equation}
    \begin{equation}
D_{\alpha \beta}[1] = \gamma_{\alpha \beta}\frac{N-3}{6(N-1)}
\Gamma \left(3 - \frac{N}{2} \right)
\left( K + \frac{24 \, \alpha_N}{(N-2)(N-3)(N-4)} \right),
\label{D1ab}
\end{equation}
        \begin{eqnarray}
D_{00}[2] = \frac{N\!-\!2}{a^2 \, 36} \left\{ \left[
\Gamma \left(4 \!-\! \frac{N}{2} \right) (N\!-\!4) c^2 +
\Delta \xi \, \Gamma \left(3 \!-\! \frac{N}{2} \right) 6(N\!-\!1)
\right. \right. \times
        \nonumber    \\
\times \, \biggl( (N-2) K + (N-6) c^2 \biggr) \left]
\left[ K + \frac{24 \, \alpha_N}{(N\!-\!2)(N\!-\!3)(N\!-\!4)} \right]
\right. +           \nonumber    \\
+ \left.\Gamma \left(4 \!-\! \frac{N}{2} \right)
\left( \frac{(5N\!-\!8)K^2}{10} - \frac{576 \, \beta_N}
{(N\!-\!2)(N\!-\!3)(N\!-\!4)(N\!-\!5)(N\!-\!6)}
\right) \right\},
\label{D200}
\end{eqnarray}
        \begin{eqnarray}
D_{\alpha \beta}[2] = \frac{\gamma_{\alpha \beta} (N-2)}{a^2\, 36 (N-1)}
 \left\{ \left[ \Gamma \left(4 - \frac{N}{2} \right) (N\!-\!4)
\biggl( (5\!-\!N) c^2 - 2c' \biggr) \right. \right. +
        \hspace{-4mm}  \nonumber    \\
+\left.\Delta \xi  \Gamma \!\left(3 \!-\! \frac{N}{2} \right) 6(N\!-\!1)
\biggl( (5\!-\!N) (N\!-\!2) K + (N\!-\!6) \left( (5\!-\!N) c^2 \!-\!
2c'\right) \biggr) \right]  \! \times
\hspace{-4mm}   \nonumber     \\[1mm]
\times \left[ K + \frac{24 \, \alpha_N}{(N\!-\!2)(N\!-\!3)
(N\!-\!4)} \right] +
 \Gamma \left(4 - \frac{N}{2} \right) \frac{(5\!-\!N)(5N\!-\!8)}{10}
\times          \hspace{-4mm}  \nonumber    \\
\times \left.   \left( K^2 - \frac{5760 \, \beta_N}
{(N\!-\!2)(N\!-\!3)(N\!-\!4)(N\!-\!5)(N\!-\!6)(5N\!-\!8)}
\right) \right\}. \phantom{xxx}
\label{D2ab}
\end{eqnarray}
    Therefore, for the geometric structure of the first and the second
subtractions in the $n$-wave procedure and in the effective action method
to coincide, it is necessery that the same
equalities~(\ref{flaN}), (\ref{flbN}) that are obtained in Appendix~A
for integer values of $N$ and $K=0,1$ be satisfied for arbitrary~$N$.
    For $N=4$, we have $\alpha_N=\beta_N=0$, which agrees with~(\ref{mera}).
    Thus, there exists a continuation with respect to the dimension of
the integration measure in the momentum space~$\{{\bf \lambda} \}$
such that the geometric structures of all the subtractions in
the $n$-wave procedure for the four-dimensional homogeneous isotropic
space-time and in the effective action method coincide.

     The evaluated counterterms have a purely geometric structure
(see~(\ref{TE0}), (\ref{TE1}), and (\ref{T2ikc}))
and can be represented as variational derivatives of geometric
quantities with respect to the metric.
    Therefore, a Lagrangian for the gravitational field can be constructed
such that adding these counterterms leads to a renormalization of its
parameters
(see a discussion of this approach in Sec.~6.11 in~\cite{DeWitt75}
and also in Sec.~6.1 in~\cite{BD}).
    Because the counterterms contain expressions that are quadratic in
the curvature, such a bare gravitational Lagrangian obviously does not
coincide with the standard Einstein Lagrangian.
    The existence of expressions quadratic in the curvature indicates
that quantization of fields in curved space inevitably involves
going beyond the Einstein theory of gravity~(Sec.~8.4 in~\cite{GMM})
    The values of the renormalized constants must be determined
experimentally, and as indicated, e.g., in~\cite{GMM},
it is possible that the renormalized constants at the quadratic
terms are equal to zero.

    Using formulas~(\ref{TE0}), (\ref{TE1}), (\ref{T2ikc}), and
       \begin{equation}
R^{\,lmpq} R_{lmpq} - 4 R^{\,lm} R_{\,lm} + R^2 =   C_{lmpq} C^{lmpq}
- \frac{4(N\!-\!3)}{N\!-\!2} \biggl( R_{lm} R^{\,lm} -
\frac{N\,R^2}{4(N\!-\!1)} \biggr),
\label{RRRRRC}
\end{equation}
     we can draw the next conclusion.
     The first three $n$-wave procedure subtractions in the
$N$-dimensional homogeneous isotropic space-time correspond to
a renormalization of the cosmological and gravitational constants and
of the parameters at the terms in the bare gravitational Lagrangian
that are quadratic in the curvature and have the form
     \begin{equation}
L_{gr,\,\varepsilon}=\sqrt{|g|} \left[ \frac{1}{16 \pi G_\varepsilon} \,
( R -2 \Lambda_\varepsilon ) + \alpha_\varepsilon \, \biggl(
R^{ik} R_{ik} - \frac{N \, R^2}{4(N-1)}  \, \biggr) +
\beta_\varepsilon \, R^2 \right].
\label{Lgr0}
\end{equation}
    For the conformal scalar field $(\xi=\xi_c)$, the parameters
$G_\varepsilon$ and $\beta_\varepsilon$ have a finite renormalization
as $N \to 4$  (see~(\ref{TE1}) and (\ref{T2ikc})).
    But as $N \to 4$, the renormalization of the parameter
$\alpha_\varepsilon $
in accordance with  (\ref{TE2ik}), (\ref{Eik}), and (\ref{RRRRRC})
is infinite $(\sim (N-4)^{-1})$  for any value of the coupling
constant $\xi$ of the scalar field to the curvature.

\section{Conclusions}
\hspace{\parindent}
    We have investigated the geometric structure of counterterms
of the vacuum EMT of a scalar field with an arbitrary coupling to
the curvature in the $N$-dimensional homogeneous isotropic spase-time.
    The obtained formulas (\ref{Ttik})--(\ref{tsab}) determine
nonrenormalized expectations of the EMT of a complex scalar field
with an arbitrary coupling to the curvature, taken in the vacuum
determined by the Hamiltonian diagonalization method~\cite{Pv}.
   The geometric structure of counterterms were determined in the
effective action method for the $N$-dimensional homogeneous isotropic
space (see~(\ref{TE0}), (\ref{TE1}), and (\ref{T2ikc})).
    We have generalized the $n$-wave procedure to homogeneous
isotropic $N$-dimensional spaces and determined the corresponding
counterterms~(\ref{at00})--(\ref{tab2}).
    We have found the properties (see~(\ref{gdfF})--(\ref{nfF}))
of eigenfunctions of the Laplace-Beltrami operator in higher-dimensional
homogeneous isotropic spaces, which are necessary for calculating
the vacuum EMT.
    The geometric structure of counterterms in the $n$-wave procedure
has been analyzed using dimensional regularization.
    We have found an analitic continuation with respect to the dimension
of the integration measure in the momentum space
(see~(\ref{sig}), (\ref{flaN}), and (\ref{flbN})).
    We have shown the coincidence of the geometric structures of
the first three subtraction in the $n$-wave procedure and in
the effective action method.
    We have shown that the first three subtractions (which exhaust all
subtractions in dimensions $N=4,5$) of the $n$-wave procedure in
the $N$-dimensional homogeneous isotropic space-time correspond to
renormalization of the cosmological and gravitational constants and
of the parameters at the terms in the bare gravitational Lagrangian
that are quadratic in the curvature~(\ref{Lgr0}).

    Comparison of the $n$-wave procedure
subtractions~(\ref{at00})--(\ref{tab2}) for $N=4$ with the counterterms
in the adiabatic regularization method, given in~\cite{Bunch80},
demonstrates the equivalence of these methods for eliminating divergences.
    Therefore, conclusions regarding the geometric structure of
subtractions obtained in this work for the $n$-wave procedure
in a four-dimensional homogeneous isotropic space-time
are also valid in the adiabatic regularization method.

\vspace{7mm}
\noindent
{\bf \Large        Appendix\ A  }

\setcounter{equation}{0}
\renewcommand{\theequation}{A.\arabic{equation}}

\vspace{2mm}
    We give some properties of the complete orthonormal set of
the eigenfunctions $\Phi_J({\bf x})$ of the Laplace-Beltrami
operator $\Delta_{N-1}$ in an $(N-1)$-dimensional space with the metric
$d l^2=\gamma_{\alpha \beta}\, d x^\alpha d x^\beta $
of constant curvature $K=0, \pm 1$.
    These properties are used to calculate the vacuum expectations
of the EMT.
    We set
     \begin{equation}
f(\lambda) = \hspace{-25pt} \sum_{\phantom{xxxxx} J\, {\scriptscriptstyle
(\lambda = {\rm const})}} \hspace{-17pt} |\Phi_J({\bf x})|^2   \,,
\label{fF}
\end{equation}
    where $\lambda$ is defined in (\ref{DFlF}) and where summation
is replaced with integration for continuous $J$.
    In the case of a homogeneous isotropic space, the function
$f(\lambda)$ is independent of $\bf x $.\
    Applying the operator $\Delta_{N-1}$ to (\ref{fF}) and
taking~(\ref{DFlF}) into account, we obtain
    \begin{equation}
\sum_{\phantom{xxxxx} J\, {\scriptscriptstyle (\lambda={\rm const})}}
\hspace{-17pt} \gamma^{\alpha \beta }
\partial_\alpha \Phi^*_J({\bf x})\,
\partial_\beta \Phi_J({\bf x}) =
\biggl( \lambda^2 - \biggl(\frac{N-2}{2} \biggr)^2 \, K \biggr)\,
f(\lambda)  \,.
\label{gdfF}
\end{equation}
    From the space isotropy condition with~(\ref{gdfF}) taken into
account, we obtain
     \begin{equation}
            \sum_{\phantom{xxxxx} J\,
{\scriptscriptstyle (\lambda={\rm const})}} \hspace{-17pt}
\partial_\alpha \Phi^*_J(x)\, \partial_\beta \Phi_J(x) =
\frac{ \gamma_{\alpha \beta }}{N-1} \, \biggl( \lambda^2 -
\biggl(\frac{N-2}{2} \biggr)^2 \, K \biggr)\, f(\lambda)  \,.
\label{dfF}
\end{equation}
     Applying the covariant derivatives
$\tilde{\nabla}_{\!\alpha}\, \tilde{\nabla}_{\!\beta}$
in the $(N-1)$-dimensional space
to (\ref{fF}) and recalling (\ref{dfF}), we obtain
    \begin{equation}
\hspace*{-25pt} \sum_{\phantom{xxxx} J\,
{\scriptscriptstyle (\lambda={\rm const})}}
\hspace{-22pt} \left[\left(\tilde{\nabla}_{\!\alpha}
\tilde{\nabla}_{\!\beta} \Phi^*_J \right) \Phi_J +
\Phi^*_J \tilde{\nabla}_{\!\alpha} \tilde{\nabla}_{\!\beta} \Phi_J \right]
= - \, \frac{2 \gamma_{\alpha \beta }}{N\!-\!1} \,
\biggl( \lambda^2\! -\! \biggl(\frac{N\!-\!2}{2} \biggr)^2 K \biggr)\,
f(\lambda) \,.
\label{nfF}
\end{equation}
    In the quasi-Euclidean case ($K=0$) in Cartesian coordinates,
\begin{equation}
\Phi_J({\bf x}) = (2 \pi)^{-(N-1)/2} \exp (-i \lambda_\alpha x^\alpha)\,,
\label{Phi}
\end{equation}
    where $\ \ -\infty < \lambda_\alpha < +\infty \ $.
    Integrating over the sphere of radious $\lambda$ in the
$(N-1)$-dimensional space of
``dimensionless momenta'' $\{ \lambda_\alpha \}$ gives
    \begin{equation}
f(\lambda)=\frac{B_N}{2}\,\lambda^{N-2} \,, \hspace{37pt} K=0 \,,
\label{flK0}
\end{equation}
     where $ B_N $ is defined in Sec.~3.

    In the case where $K=1$ (spherical space) for a fixed $\lambda$,
the set of the indices $J$ is finite, and the volum of the
$(N-1)$-dimensional space is given by $S_N(1)$, the surface area of
the unit sphere in the $N$-dimensional space.  Therefore,
$$
f(\lambda)=\frac{1}{S_N(1)} \int  \biggl( \hspace{-25pt}
\sum_{\phantom{xxxxx} J\, {\scriptscriptstyle (\lambda={\rm const})}}
\hspace{-21pt} \Phi^*_J({\bf x}) \Phi_J({\bf x})  \biggr)
\sqrt{\gamma}\, d^{N-1}{\bf x} \,,
$$
   where $ \gamma\!=\!{\rm det}(\gamma_{\alpha \beta})$.
     Changing the order of the summation and integration operations
and recalling that $\Phi_J({\bf x})$ are orthonormalized, we obtain
    \begin{equation}
f(\lambda)=\dim \lambda\, / S_N(1) \ ,
\label{dfS}
\end{equation}
    where $ \dim \lambda \, $ is the multiplicity of the eigenvalue
$\lambda^2-((N\!-\!2)/2)^2$ of the operator $-\Delta_{N-1}$.
    Using the relation expressing the multiplicity of the eigenvalues
of the Laplace-Beltrami operator on the sphere through the dimension
of the space of harmonic polynomials
(see, e.g., \cite{Shu}), we obtain the formula
    \begin{equation}
f(\lambda)=\frac{B_N}{2} \, \frac{\Gamma (n\!+\!N\!-\!2)}{\Gamma (n+1)}\,
\lambda\,,
\ \ \ n=0,1, \ldots \,, \ \ \ \lambda=n+\frac{N-2}{2} \,.
\label{flK1}
\end{equation}
    For an arbitrary $N$ and $\ K=0,1$,
expressions~(\ref{flK0}) and (\ref{flK1}) can be written as
    \begin{equation}
f(\lambda)=\frac{B_N}{2} \left(\lambda^{N-2} + \alpha_N \lambda^{N-4} +
\beta_N \lambda^{N-6} + \ldots \right),
\label{flKab}
\end{equation}
    where
    \begin{equation}
\alpha_N = - \frac{1}{24}\,(N-2) (N-3) (N-4)\, K \,,
\label{flaN}
\end{equation}
    \begin{equation}
\beta_N = \frac{1}{5760}\,(N-2) (N-3) (N-4) (N-5) (N-6) (5N-8)\, K^2 \,.
\label{flbN}
\end{equation}

\vspace{4mm}
\noindent
{\Large \bf Appendix\ B }

\setcounter{equation}{0}
\renewcommand{\theequation}{B.\arabic{equation}}

\vspace{2mm}
    We give expressions for some geometric quantities in the
$N$-dimensional homogeneous isotropic space-time with metric~(\ref{gik}).

      The nonzero Christoffel symbols are
     \begin{equation}
\Gamma^{\,0}_{\, 00}= \frac{a'}{a} \equiv c   \ ,
\ \ \ \  \Gamma^{\,\alpha}_{\, 0 \beta}=c \, \delta^\alpha_\beta   \ ,
\ \ \ \ \Gamma^{\,0}_{\, \alpha \beta}= c \, \gamma_{\alpha \beta} \ ,
\ \ \ \ \Gamma^{\,\alpha}_{\, \beta \delta}(g_{ik}) =
\Gamma^{\,\alpha}_{\, \beta \delta}(\gamma_{\nu \mu}) \,.
\label{GGG}
\end{equation}
    The nonzero Ricci tensor components and the scalar curvature are
     \begin{equation}
R_{00} = (N-1) \, c'   \ , \ \ \ \
R_{\alpha \beta}=-\gamma_{\alpha \beta} \left[ \, c'+(N-2) (c^2+K)
\right],
\label{R00Rab}
\end{equation}
     \begin{equation}
R= a^{-2}(N-1) \left[ \, 2c'+(N-2)(c^2+K) \right].
\label{RRRR}
\end{equation}
      The Einstein tensor components are
     \begin{equation}
G_{00} =\! -\frac{(N\!-\!1) (N\!-\!2)}{2} (c^2\!+\!K) ,  \ \
G_{\alpha \beta}=\! \gamma_{\alpha \beta} (N\!-\!2)
\left[ c'+\frac{(N\!-\!3)}{2} (c^2\!+\!K)\right].
\label{G00Gab}
\end{equation}
     Using Eqs.(\ref{GGG})--(\ref{RRRR}), we obtain
the square of the Ricci tensor and the second derivatives in the form:
     \begin{equation}
R_{\,lm} R^{\,lm} = a^{-4}\,(N-1) \left[ N c'{\,}^2 +
2 (N-2) \, c'(c^2 +K) + (N-2)^2 (c^2 +K)^2 \right],
\label{RlmRlm}
\end{equation}
       \begin{eqnarray}
\nabla_{\! 0} \nabla_{\! 0} R &=&  a^{-2} \, 2 (N-1) \left[ \, c^{(3)} +
(N-7) \, c'' c  + (N-4)\, c'{\,}^2  \right. -
        \nonumber           \\
&-& \left. 6(N-3)\, c' c^2 + 3(N-2) c^4 + K(N-2) (3c^2 -c') \right],
\label{D0D0R}
\end{eqnarray}
       \begin{equation}
\nabla_{\! \alpha} \nabla_{\! \beta} R = \gamma_{\alpha \beta} a^{-2} \,
2 (N\!-\!1) \left[ - c'' c  - (N\!-\!4) c' c^2 +  (N\!-\!2)  c^2
(c^2 + K) \right],
\label{DaDbR}
\end{equation}
       \begin{eqnarray}
\nabla^l \nabla_{\! l} R =  a^{-4} \, 2 (N-1) \left[ \, c^{(3)} +
2 (N-4) \, c'' c  + (N-4)\, c'{\,}^2  \right. +
        \nonumber           \\
+\left. ( N^2-10 N +20 )\, c' c^2 -  (N-2) ((N-4) c^2 + c')\,(c^2 + K)
\right].
\label{DDR}
\end{eqnarray}
       Using (\ref{R00Rab}), (\ref{RRRR}), (\ref{RlmRlm})--(\ref{DDR}),
we obtain the components ${}^{(1)}\! H_{ik} $ and ${}^{(3)}\! H_{ik} $
(see~(\ref{1Hik}), (\ref{3Hik})) as
       \begin{eqnarray}
{}^{(1)}\! H_{00}= \frac{(N\!-\!1)^2}{a^2} \left[ 2 c'{\,}^2 - 4 c''c
- 4 (N\!-\!4) c' c^2 - \frac{c^4}{2} (N\!-\!2) (N\!-\!10)  \right.  -
     \nonumber     \\
-\left. K(N-2) \left( \frac{N-2}{2}K + (N-6) c^2 \right)
\right], \phantom{xxxxx}
\label{c1H00}
\end{eqnarray}
       \begin{eqnarray}
{}^{(1)}\! H_{\alpha \beta} = \gamma_{\alpha \beta} \, a^{-2} (N-1) \,
\biggl\{ 4 c^{(3)} + 4\, ( 2 N - 9 )\, c''c + 2\, (3 N - 11) \, c'{\,}^2 +
\nonumber           \\
+\, 6\, ( N^2 - 10 N + 20 ) \, c' c^2 +
\frac{1}{2} ( N-2 ) ( N^2 - 15 N + 50 ) \, c^4 +
\nonumber    \\
+\, K(N\!-\!2)\, \biggl[ (N\!-\!5)(N\!-\!6) c^2 + 2(N\!-\!6) c' +
\frac{1}{2} (N\!-\!2)(N\!-\!5) K \, \biggr] \biggr\} , \phantom{xx}
\label{c1Hab}
\end{eqnarray}
     \begin{equation}
{}^{(3)}\! H_{00} = a^{-2}\,2^{-1} (N-1) (N-2) (N-3) \, (c^2 + K)^2  \ ,
\label{c3H00}
\end{equation}
     \begin{equation}
{}^{(3)}\! H_{\alpha \beta} = - \gamma_{\alpha \beta}\, a^{-2} \, 2^{-1}
(N-2) (N-3) \left[ 4 c' (c^2 +K) + (N-5) \, (c^2 +K)^2 \right].
\label{c3Hab}
\end{equation}

\vspace{4mm}
 {\bf \large Acknowledgments.}
The author is grateful to Prof. A.\,A.\,Grib for helpful discussion.
This work was supported in part by the Russian Federation Ministry of
Education (Grants Nos. E00-3-163 and E02-3.1-198).

\newpage

\end{document}